# High-performance green and blue quantum-dot light-emitting diodes with eliminated charge leakage

**Authors:** Yunzhou Deng[1,4]†, Feng Peng[2]†, Yao Lu[3]†, Xitong Zhu[1]†, Wangxiao Jin[1], Jing Qiu[3], Jiawei Dong[3], Yanlei Hao[1], Dawei Di[4], Yuan Gao[5], Tulai Sun[6], Linjun Wang[3]*, Lei Ying[2]*, Fei Huang[2]* and Yizheng Jin[1]*

**Affiliations:**

[1]Key Laboratory of Excited-State Materials of Zhejiang Province, State Key Laboratory of Silicon Materials, Department of Chemistry, Zhejiang University, Hangzhou 310027, China.

[2]Institute of Polymer Optoelectronic Materials and Devices, State Key Laboratory of Luminescent Materials and Devices, South China University of Technology, Guangzhou 510640, China.

[3]Key Laboratory of Excited-State Materials of Zhejiang Province, Department of Chemistry, Zhejiang University, Hangzhou 310027, China.

[4]State Key Laboratory of Modern Optical Instrumentation, College of Optical Science and Engineering, International Research Center for Advanced Photonics, Zhejiang University, 310027, Hangzhou, China.

[5]Najing Technology Corporation Ltd, Hangzhou 310052, China.

[6]Center for Electron Microscopy, State Key Laboratory Breeding Base of Green Chemistry Synthesis Technology and College of Chemical Engineering, Zhejiang University of Technology, Hangzhou 310014, China.

†These authors contribute equally to this work.

*Corresponding author: Prof. Linjun Wang (ljwang@zju.edu.cn), Prof. Lei Ying (msleiying@scut.edu.cn), Prof. Fei Huang (msfhuang@scut.edu.cn) or Prof. Yizheng Jin (yizhengjin@zju.edu.cn).

**Abstract:** Quantum-dot light-emitting diodes (QD-LEDs) promise a new generation of efficient, low-cost, large-area and flexible electroluminescent devices. However, the inferior performance of green and blue QD-LEDs is hindering the commercialization of QD-LEDs in display and solid-state lighting. Here, we demonstrate best-performing green and blue QD-LEDs with ~100% conversion of the injected charge carriers into emissive excitons. Key to this success is eliminating electron leakage at the organic/inorganic interface by using hole-transport polymers with low electron affinity and reduced energetic disorder. Our devices exhibit record-high peak external quantum efficiencies (28.7% for green, 21.9% for blue), exceptionally high efficiencies in wide ranges of luminance, and unprecedented stability ($T_{95}$ lifetime: 580,000 h for green, 4,400 h for blue). The overall performance surpasses previously reported solution-processed green and blue LEDs.

**Main Text:** LEDs using solution-processable emitters, such as colloidal QDs (*1-11*), conjugated polymers (*12*), and metal-halide perovskites (*13, 14*), are attractive for next-generation display and lighting technologies, owing to the potential for low-cost fabrication of efficient, large-area and flexible electroluminescent (EL) devices. Colloidal QDs are a unique class of solution-processable inorganic crystals featuring efficient, stable, and high color-purity luminescence (*15-19*). CdSe-based QDs (*1-9*), and more recently, InP-based (*10*) and ZnSe-based (*11*) QDs have been applied as the emissive materials in LEDs. Fig. 1A shows a typical structure of the state-of-the-art QD-LEDs, which consists of QDs sandwiched between electron-transport layers (ETLs) based on zinc oxide (ZnO) nanoparticles and hole-transport layers (HTLs) based on polymers. To promote the performance of QD-LEDs, several effective strategies, including advanced synthetic chemistry for enhancing the photoluminescence quantum yield (PLQY) of QDs (*10, 11, 20, 21*), band-structure tailoring of QDs for hole-injection improvement (*7, 9, 22*), the use of electrochemically-stable ligands for eliminating in-situ redox reactions (*23*), and material design of ZnO ETLs for achieving efficient electron injection and suppressed interfacial exciton quenching (*3, 5, 24*), have been developed. These efforts have enabled red QD-LEDs with high EQEs (>20%) and long operational lifetimes ($T_{95}$ at 100 cd m$^{-2}$: >300,000 h), fulfilling the requirements of EL displays (*21, 24*). However, EL efficiencies of the state-of-the-art green and blue QD-LEDs are still lower than the limits considering the near-unity PLQYs of the QDs and the light out-coupling efficiencies of LEDs. The operational lifetimes of the green (*9*) and blue (*11, 23*) devices are also far below those of the red devices. These facts call for new design strategies to develop high-performance green and blue QD-LEDs.

We investigate the discrepancies of efficiency losses in red, green, and blue-colored QD-LEDs with Cd-based QDs as emitters. Red (CdSe/CdZnSe/ZnS) core/shell/shell QDs, green (CdSe/CdZnSe/ZnS) core/shell/shell QDs, and blue (CdZnSe/ZnS) core/shell QDs with similar diameters (~10 nm) and surface ligands (Fig. S1) are incorporated into a unified QD-LED structure (Fig. S2A). Poly[(9,9-dioctylfluorenyl-2,7-diyl)-co-(4,4′-(N-(4-sec-butylphenyl)diphenylamine)] (TFB), the benchmark hole-transport material for QD-LEDs (*7, 9-11, 22*), is employed as HTLs. All three devices show sub-bandgap turn-on characteristics (Fig. 1B), i.e., the turn-on voltage being lower than the voltage corresponding to the emission photon energy ($V_{photon}$), indicating efficient charge injection in these QD-LEDs. The internal quantum efficiency (IQE) of the red QD-LED determined from the external quantum efficiency (EQE) and out-coupling efficiency (Fig. S2) is close to the limit defined by the PLQY of the red QD film (Fig. 1C). In contrast, there are pronounced EL-PL efficiency gaps—discrepancies between the IQE of the EL devices and the PLQYs of the QD films—for the green and blue QD-LEDs (Fig. 1C).

Spectral characterizations indicate the occurrence of electron leakage into the HTLs in the green and blue QD-LEDs. The red QD-LED shows pure QD emission according to the EL spectra measured at a current density of 100 mA cm$^{-2}$ or the voltage corresponding to the peak EQE (Fig. 1D). In contrast, parasitic emissions from TFB are observed in the EL spectra of the green and blue QD-LEDs (arrows in Fig. 1D, and see Fig. S3 for EL properties of TFB), suggesting exciton formation in the TFB layer. Transient PL characterizations (Fig. S4) rule out the possibilities of energy-transfer processes from QDs to TFB or exciton dissociation induced by TFB. Hence, in the operation of the blue and green QD-LEDs, exciton formation in the TFB layer suggests electron leakage, i.e., electron transfer from the negatively charged QDs to the TFB ( QD$^-$+HTL→QD+HTL$^-$ ). Given the low EL efficiency of TFB (~0.01%, Fig. S3), the distinguishable parasitic emission (intensity: ~3 orders of magnitude lower than that of QD emission) in EL spectra of the green and blue QD-LEDs indicate non-negligible electron leakage in the device operation, causing the EL-PL efficiency gaps.

It is of interest to understand the underlying mechanism that enables the interfacial electron leakage between the QDs and the HTLs. A glance at the energy-level diagram at the QD/HTL interface (Fig. S5) would infer that the large energy offset between the lowest unoccupied molecular orbital (LUMO) of TFB and the conduction band (CB) of QDs ($\Delta E_{\text{LUMO,HTL–CB,QD}} > 0.5$ eV) shall make the electron transfer from QDs to TFB extremely inefficient, if not impossible. We consider that this conventional picture is oversimplified because the interfacial charge transfer rate also relies on the densities of initial and final electronic states involved in the transfer process. Given the uniqueness of the interface between the crystalline inorganic QDs and the amorphous organic HTLs (Fig. 1A, right), we hypothesize that the interfacial electron transfer can be enhanced by two material-specific factors—the energetic disorder of the HTLs and the size discrepancy between the QDs and the HTL segments. For the polymeric HTLs, both static disorder (denoted by $\sigma$, the Gaussian standard derivation of density-of-state, DOS, distribution in the frontier orbitals) and dynamic disorder (denoted by $\lambda$, the reorganization energy) contribute to the energetic disorder near the LUMO (Fig. 1E). The former is due to variations in conformation, conjugation length of the segments (*25, 26*), or defects and the latter originates from the strong electron-phonon interactions (*27, 28*). Furthermore, we emphasize that a QD crystal (diameter: ~10 nm) is significantly larger than an HTL segment (size: ~1–2 nm), leading to that the electron in a QD donor can be transferred to multiple HTL acceptors (Fig. 1A).

We carry out mixed quantum-classical simulations of the electron-transfer dynamics in one-dimensional (1D), two-dimensional (2D), and three-dimensional (3D) interfacial models (Fig. 1F), which consider the effects of $\sigma$, $\lambda$ and $\Delta E_{\text{LUMO,HTL–CB,QD}}$. Two QDs on the right side are coupled with one, four, or sixteen HTL chains on the left side in 1D, 2D, and 3D cases, respectively, and each HTL chain consists of ten effective conjugated units (*29*). The electron-transfer dynamics from QDs to HTLs is simulated with the state-of-the-art crossing corrected global flux surface hopping method implemented in a homemade SPADE simulation package, which is a robust tool for simulating general nonadiabatic dynamics in extended systems (Supplementary Text for details) (*30-32*).

Our theoretical investigation suggests a unique mechanism for electron transfer at the QD/HTL interface (Fig. S6), which is distinctive from those at the inorganic/inorganic interfaces or the organic/organic interfaces. Namely, the dynamics are generally initialized by several incoherent electron hops or an indirect super-exchange process, driven by both energetic disorders of the polymer HTLs and the entropy gain originated from the different numbers of QD donors and HTL acceptors. The transferred electron is further relaxed by the strong structural reorganization in the HTLs, thus preventing the back transfer of the stabilized electron from HTLs to the rigid QDs (see Fig. S6 for representative trajectories).

The electron-leakage probability was simulated with different sets of parameters. The results (Fig. 1G) show that in the absence of static disorder and size difference between QDs and HTLs (grey-open triangles), the electron-transfer probability for a small $\lambda$ across a large energy barrier ($\Delta E_{\text{LUMO,HTL–CB,QD}} > 0.5$ eV) is negligible (< 0.02%), which is consistent with the conventional understanding. With the increased $\lambda$ of the HTLs, the electron transfer from QDs to HTLs for a given $\Delta E_{\text{LUMO,HTL–CB,QD}}$ is enhanced in the 1D model (grey-open circles). Notably, when the static energetic disorder of the HTL segments is introduced, the electron-transfer probability is enhanced by several orders of magnitude for a large $\Delta E_{\text{LUMO,HTL–CB,QD}}$ of over 0.5 eV (grey-solid circles). Moreover, the results of the 2D and 3D configurations indicate a further increased electron-transfer probability (red and blue solid circles). More details are shown in Fig. S7. Overall, the results predict a non-negligible electron-transfer probability despite a seemly huge $\Delta E_{\text{LUMO,HTL–CB,QD}}$, and support the existence of the electron-leakage channels in the blue and green QD-LEDs (Fig. 1E). Regarding the red QD-LEDs, the absence of such an electron-leakage channel is reasonable because of the relatively small upshift of the CBs of red QDs with respect to that of bulk CdSe and hence, a considerably larger $\Delta E_{\text{LUMO,HTL–CB,QD}}$ of > 1.0 eV (Fig. S5).

These findings motivate us to propose a new design principle for the QD/HTL interface to boost the performance of green and blue QD-LEDs. As shown in Fig. 2A, a conventional strategy is to enhance hole injection for more efficient exciton formation (QD$^-$+HTL$^+$→QD$^X$+HTL) (*7, 9*). Here, we propose to use HTLs with shallower LUMO and reduced energetic disorders to suppress the electron-leakage channel (QD$^-$+HTL→QD+HTL$^-$).

Our new principle requires the molecular structure of HTLs to simultaneously possess rigid backbones and limited conjugation. Accordingly, we employ a co-polymer of poly[(9,9-dioctylfluorenyl-2,7-diyl)-alt-(9-(2-ethylhexyl)-carbazole-3,6-diyl)] (PF8Cz). Compared with the benchmark HTL of TFB (Fig. 2B, left), the non-planar, propeller-like triphenylamine unit (*33*) is replaced by a planar 3,6-carbazole unit (Fig. 2B, right), providing a more rigid framework and thus reducing the energetic disorder. Furthermore, for 3,6-carbazole-based copolymers, the nitrogen atom possesses strong electron-donating properties and meanwhile breaks the conjugation (*34, 35*). Fig. 2B shows the theoretically optimized geometrics of PF8Cz and TFB dimers (Fig. S8 for details). Density-functional theory (DFT) calculations demonstrate considerably smaller reorganization energy of PF8Cz ($\lambda$: 0.32 eV for monomers and 0.21 eV for dimers) than TFB ($\lambda$: 0.47 eV for monomers and 0.28 eV for dimers), indicating reduced electron-phonon interactions and dynamic disorder (Fig. S9 and Table S1 for details). Furthermore, PF8Cz possesses a higher LUMO energy relative to that of TFB (by 0.24 eV) because of the more localized electron distribution on the fluorene unit (Fig. S8).

The desired electronic properties of PF8Cz (Fig. 2B) are validated by spectroscopic characterizations. Both the absorption and PL spectra of the PF8Cz film are blue-shifted compared with those of the TFB film (Fig. 2C). The optical bandgap of the PF8Cz film is 3.06 eV, larger than that of the TFB film, 2.88 eV. UPS measurements indicate an identical highest occupied molecular orbital (HOMO) onset energy (−5.4 eV) for both the PF8Cz and TFB films (Fig. 2D). Accordingly, the LUMO level of the PF8Cz film is estimated to be upshifted by ~0.2 eV compared with that of the TFB film (Fig. 2B), which favors the blocking of the electron-transfer channel. Furthermore, the DOS profiles and the band-bending analyses verify the reduced energetic disorder of the PF8Cz film. The widths of the high-kinetic-energy peaks of valence spectra correspond to the extent of DOS broadening caused by energetic disorder (*36, 37*). According to Fig. 2D, Gaussian fitting obtains a HOMO DOS width ($\sigma$) of 0.25 ± 0.02 eV for the PF8Cz film, narrower than that of the TFB film, 0.37 ± 0.04 eV. This result is in line with our analyses on the band-bending profiles (Fig. 2E), i.e., shifts in the surface work functions (or vacuum levels) with increased thicknesses of PF8Cz or TFB films on conductive PEDOT:PSS substrates. Given the similar HOMO levels of TFB and PF8Cz films, the degree of band bending, which is ascribed to spontaneous charge transfer from the substrates into the tail states of the polymers, can be correlated with the energetic disorder of the HTLs (*26, 37, 38*). The smaller shifts in the work function of the PF8Cz films on PEDOT:PSS than those of the TFB films on PEDOT:PSS (Fig. 2E) imply a narrower distribution of HOMO DOS of PF8Cz.

We evaluate the electron-blocking properties of HTLs by analyzing band-bending characteristics of the HTL films deposited onto low-work-function substrates of Sm, which results from the spontaneous electron transfer from Sm to HTLs. TFB films on Sm show pronounced band bending while the surface work functions of the PF8Cz films on Sm are almost constant (Fig. 2E). The results verify that the higher LUMO level and reduced energetic disorder of the PF8Cz films than those of the TFB films lead to superior electron-blocking properties of PF8Cz.

Finally, we show that by using PF8Cz as the HTL, substantial improvements in EL performance are achieved in both green and blue QD-LEDs. The green and blue devices exhibit sub-bandgap turn-on voltages (Fig. 3A). Notably, the parasitic emission of HTLs is substantially suppressed in the PF8Cz-based QD-LEDs when compared with that of the TFB-based devices (Fig. 3B, and see Fig. S10A for EL properties of PF8Cz). This feature validates the desirable electron-blocking properties of PF8Cz enabled by effective modulation of LUMO level and energetic disorder. Consequently, the PF8Cz-based QD-LEDs demonstrate unprecedentedly high efficiencies. A power efficiency of 162 lm $W^{-1}$ and a current efficiency of 127 cd $A^{-1}$ are realized for the green device (Fig. 3C). The green and blue QD-LEDs show peak EQEs of 28.7% and 21.9%, respectively (Fig. 3D), both representing the highest values for QD-LEDs. Peak IQEs of the green (~94%) and blue QD-LEDs (~86%) have approached the limits defined by the PLQYs of the QD films, suggesting the elimination of the EL-PL efficiency gap (Fig. 3E). Current-density-voltage characteristics of hole-only devices (Fig. S10B) and the QD-LEDs (Fig. 1B and Fig. 3A) show no evidence of enhanced hole transport in the PF8Cz-based QD-LEDs and further validate that the efficiency improvement is primarily due to the enhanced electron-blocking properties of the PF8Cz HTLs. Impressively, the green QD-LEDs show EQEs of >20% for luminance across four orders of magnitude (20 cd $m^{-2}$ to over 200,000 cd $m^{-2}$), and the EQEs of the blue QD-LEDs maintain to be >20% in a wide luminance range of 400–25,000 cd $m^{-2}$. These wide high-efficiency windows are exceptional among solution-processed LEDs (Fig. S12) and readily cover the luminance required for flat-panel display and general lighting applications (Fig. 3D).

We highlight that long operational lifetimes are achieved in the PF8Cz-based green and blue QD-LEDs. The $T_{95}$ lifetime of a green device is determined to be 71 h at an initial luminance ($L_0$) of 11,220 cd $m^{-2}$ (Fig. 4A), corresponding to a $T_{95}$ lifetime of ~580,000 h at an $L_0$ of 100 cd $m^{-2}$ (empirical acceleration factor: n = 1.91, inset of Fig. 4A and Fig. S11). This unprecedented lifetime of the green device sufficiently exceeding the requirements for EL displays. The optimized blue QD-LED realizes a $T_{95}$ lifetime of 44 h at an $L_0$ of 1,150 cd $m^{-2}$ (Fig. 4B), indicating a record-long $T_{95}$ lifetime of ~4,400 h at an $L_0$ of 100 cd $m^{-2}$ (n = 1.89, inset of Fig. 4B and Fig. S11). The $T_{50}$ lifetimes ($L_0$: 100 cd $m^{-2}$) are estimated to be ~2,570,000 h for the green QD-LEDs and ~ 24,000 h for the blue QD-LEDs (Fig. S11). According to a comparison of several key device metrics (Table S2), the overall performance of our green and blue devices surpasses all previously reported solution-processed LEDs, including QD-LEDs, organic LEDs, and perovskite LEDs.

Comparative studies were conducted on a random copolymer with the backbone consisting of 2,7-fluorene and 3,6-carbazole (denoted as R-PF8Cz), in which additional energetic disorder is introduced by the random variations in the electronic structures of the conjugation units. The results (Fig. S13) confirm the general and critical impacts of energetic disorder and electron-blocking properties of HTLs on the device performance.

In summary, we have identified a major efficiency-loss channel in green and blue QD-LEDs, i.e., the electron transfer from QDs to HTLs, which is greatly enhanced by the energetic disorder of polymeric HTLs and geometric factors of the interfacial materials. Accordingly, we offer a new design principle of employing HTLs with a shallower LUMO level and reduced energetic disorder to eliminate electron leakage. The advances in fundamental understanding close the long-existing EL-PL efficiency gap for green and blue QD-LEDs, resulting in ~100% conversion of the injected charge carriers into emissive excitons. Our devices demonstrate exceptionally high efficiencies in wide ranges of luminance and record-long operational lifetimes, representing the best-performing solution-processed green and blue LEDs. The results promise the realization of full-color QD-LEDs for next-generation display and lighting technologies. Future efforts on molecular design and synthetic strategies of the carbazole-based polymers could enhance the conductivity and electrochemical stability of the HTLs (*39*), which shall further improve the power efficiencies and operational lifetimes of green and blue QD-LEDs. Our approach for the elimination of the charge-leakage channel may inspire the design of other solution-processed LEDs with organic/inorganic interfaces.

**References and Notes**

1. V. L. Colvin, M. C. Schlamp, A. P. Alivisatos, Light-emitting-diodes made from cadmium selenide nanocrystals and a semiconducting polymer. *Nature* **370**, 354–357 (1994).
2. S. Coe, W. K. Woo, M. Bawendi, V. Bulović, Electroluminescence from single monolayers of nanocrystals in molecular organic devices. *Nature* **420**, 800–803 (2002).
3. L. Qian, Y. Zheng, J. Xue, P. H. Holloway, Stable and efficient quantum-dot light-emitting diodes based on solution-processed multilayer structures. *Nat. Photon.* **5**, 543–548 (2011).
4. J. Kwak, W. K. Bae, D. Lee, I. Park, J. Lim, M. Park, H. Cho, H. Woo, D. Y. Yoon, K. Char, S. Lee, C. Lee, Bright and efficient full-color colloidal quantum dot light-emitting diodes using an inverted device structure. *Nano Lett.* **12**, 2362–2366 (2012).
5. B. S. Mashford, M. Stevenson, Z. Popovic, C. Hamilton, Z. Zhou, C. Breen, J. Steckel, V. Bulović, M. Bawendi, S. Coe-Sullivan, P. T. Kazlas, High-efficiency quantum-dot light-emitting devices with enhanced charge injection. *Nat. Photon.* **7**, 407–412 (2013).
6. X. Dai, Z. Zhang, Y. Jin, Y. Niu, H. Cao, X. Liang, L. Chen, J. Wang, X. Peng, Solution-processed, high-performance light-emitting diodes based on quantum dots. *Nature* **515**, 96–99 (2014).
7. Y. Yang, Y. Zheng, W. Cao, A. Titov, J. Hyvonen, J. R. Manders, J. Xue, P. H. Holloway, L. Qian, High-efficiency light-emitting devices based on quantum dots with tailored nanostructures. *Nat. Photon.* **9**, 259–266 (2015).
8. X. Li, Y.-B. Zhao, F. Fan, L. Levina, M. Liu, R. Quintero-Bermudez, X. Gong, L. N. Quan, J. Fan, Z. Yang, S. Hoogland, O. Voznyy, Z.-H. Lu, E. H. Sargent, Bright colloidal quantum dot light-emitting diodes enabled by efficient chlorination. *Nat. Photon.* **12**, 159–164 (2018).
9. H. Shen, Q. Gao, Y. Zhang, Y. Lin, Q. Lin, Z. Li, L. Chen, Z. Zeng, X. Li, Y. Jia, S. Wang, Z. Du, L. S. Li, Z. Zhang, Visible quantum dot light-emitting diodes with simultaneous high brightness and efficiency. *Nat. Photon.* **13**, 192–197 (2019).
10. Y.-H. Won, O. Cho, T. Kim, D.-Y. Chung, T. Kim, H. Chung, H. Jang, J. Lee, D. Kim, E. Jang, Highly efficient and stable InP/ZnSe/ZnS quantum dot light-emitting diodes. *Nature* **575**, 634–638 (2019).
11. T. Kim, K.-H. Kim, S. Kim, S.-M. Choi, H. Jang, H.-K. Seo, H. Lee, D.-Y. Chung, E. Jang, Efficient and stable blue quantum dot light-emitting diode. *Nature* **586**, 385–389 (2020).
12. R. H. Friend, R. W. Gymer, A. B. Holmes, J. H. Burroughes, R. N. Marks, C. Taliani, D. D. C. Bradley, D. A. Dos Santos, J. L. Bredas, M. Logdlund, W. R. Salaneck, Electroluminescence in conjugated polymers. *Nature* **397**, 121–128 (1999).
13. Z.-K. Tan, R. S. Moghaddam, M. L. Lai, P. Docampo, R. Higler, F. Deschler, M. Price, A. Sadhanala, L. M. Pazos, D. Credgington, F. Hanusch, T. Bein, H. J. Snaith, R. H. Friend, Bright light-emitting diodes based on organometal halide perovskite. *Nat. Nanotechnol.* **9**, 687–692 (2014).
14. M. Yuan, L. N. Quan, R. Comin, G. Walters, R. Sabatini, O. Voznyy, S. Hoogland, Y. Zhao, E. M. Beauregard, P. Kanjanaboos, Z. Lu, D. H. Kim, E. H. Sargent, Perovskite energy funnels for efficient light-emitting diodes. *Nat. Nanotechnol.* **11**, 872–877 (2016).
15. A. P. Alivisatos, Semiconductor clusters, nanocrystals, and quantum dots. *Science* **271**, 933–937 (1996).
16. D. V. Talapin, J.-S. Lee, M. V. Kovalenko, E. V. Shevchenko, Prospects of colloidal nanocrystals for electronic and optoelectronic applications. *Chem. Rev.* **110**, 389–458 (2010).
17. C. R. Kagan, E. Lifshitz, E. H. Sargent, D. V. Talapin, Building devices from colloidal quantum dots. *Science* **353**, aac5523 (2016).

18. C. Pu, H. Qin, Y. Gao, J. Zhou, P. Wang, X. Peng, Synthetic Control of Exciton Behavior in Colloidal Quantum Dots. *J. Am. Chem. Soc.* **139**, 3302–3311 (2017).
19. F. P. G. d. Arquer, D. V. Talapin, V. I. Klimov, Y. Arakawa, M. Bayer, E. H. Sargent, Semiconductor quantum dots: Technological progress and future challenges. *Science* **373**, eaaz8541 (2021).
20. J. Lim, Y.-S. Park, K. Wu, H. J. Yun, V. I. Klimov, Droop-free colloidal quantum dot light-emitting diodes. *Nano Lett.* **18**, 6645–6653 (2018).
21. T. Lee, B. J. Kim, H. Lee, D. Hahm, W. K. Bae, J. Lim, J. Kwak, Bright and stable quantum dot light-emitting diodes. *Adv. Mater.* **2021**, 2106276.
22. W. Cao, C. Xiang, Y. Yang, Q. Chen, L. Chen, X. Yan, L. Qian, Highly stable QLEDs with improved hole injection via quantum dot structure tailoring. *Nat. Commun.* **9**, 2608 (2018).
23. C. Pu, X. Dai, Y. Shu, M. Zhu, Y. Deng, Y. Jin, X. Peng, Electrochemically-stable ligands bridge the photoluminescence-electroluminescence gap of quantum dots. *Nat. Commun.* **11**, 937 (2020).
24. D. Chen, D. Chen, X. Dai, Z. Zhang, J. Lin, Y. Deng, Y. Hao, C. Zhang, H. Zhu, F. Gao, Y. Jin, Shelf-stable quantum-dot light-emitting diodes with high operational performance. *Adv. Mater.* **32**, 2006178 (2020).
25. H. Bässler, Charge transport in disordered organic photoconductors a monte carlo simulation study. *Phys. Status Solidi B* **175**, 15–56 (1993).
26. A. Karki, G.-J. A. H. Wetzelaer, G. N. M. Reddy, V. Nádaždy, M. Seifrid, F. Schauer, G. C. Bazan, B. F. Chmelka, P. W. M. Blom, T.-Q. Nguyen, Unifying energetic disorder from charge transport and band bending in organic semiconductors. *Adv. Funct. Mater.* **29**, 1901109 (2019).
27. A. Troisi, Charge transport in high mobility molecular semiconductors: Classical models and new theories. *Chem. Soc. Rev.* **40**, 2347–2358 (2011).
28. L. Wang, Q. Li, Z. Shuai, L. Chen, Q. Shi, Multiscale study of charge mobility of organic semiconductor with dynamic disorders. *Phys. Chem. Chem. Phys.* **12**, 3309–3314 (2010).
29. S. Prodhan, J. Qiu, M. Ricci, O. M. Roscioni, L. Wang, D. Beljonne, Design rules to maximize charge-carrier mobility along conjugated polymer chains. *J. Phys. Chem. Lett.* **11**, 6519–6525 (2020).
30. J. Qiu, X. Bai, L. Wang, Crossing classified and corrected fewest switches surface hopping. *J. Phys. Chem. Lett.* **9**, 4319–4325 (2018).
31. J. Qiu, X. Bai, L. Wang, Subspace surface hopping with size-independent dynamics. *J. Phys. Chem. Lett.* **10**, 637–644 (2019).
32. L. Wang, J. Qiu, X. Bai, J. Xu, Surface hopping methods for nonadiabatic dynamics in extended systems. *WIREs Comput. Mol. Sci.* **10**, e1435 (2020).
33. J. C. Sancho-García, C. L. Foden, I. Grizzi, G. Greczynski, M. P. de Jong, W. R. Salaneck, J. L. Brédas, J. Cornil, Joint theoretical and experimental characterization of the structural and electronic properties of poly(dioctylfluorene-alt-N-butylphenyl diphenylamine). *J. Phys. Chem. B* **108**, 5594–5599 (2004).
34. P.-L. T. Boudreault, S. Beaupré, M. Leclerc, Polycarbazoles for plastic electronics. *Polym. Chem.* **1**, 127–136 (2010).
35. J. Kim, Y. S. Kwon, W. S. Shin, S.-J. Moon, T. Park, Carbazole-based copolymers: effects of conjugation breaks and steric hindrance. *Macromolecules* **44**, 1909-1919 (2011).
36. J. Hwang, A. Wan, A. Kahn, Energetics of metal–organic interfaces: New experiments and assessment of the field. *Mater. Sci. Eng. R Rep.* **64**, 1–31 (2009).
37. I. Lange, J. C. Blakesley, J. Frisch, A. Vollmer, N. Koch, D. Neher, Band bending in conjugated polymer layers. *Phys. Rev. Lett.* **106**, 216402 (2011).

38. J. C. Blakesley, N. C. Greenham, Charge transfer at polymer-electrode interfaces: The effect of energetic disorder and thermal injection on band bending and open-circuit voltage. *J. Appl. Phys.* **106**, 034507 (2009).
39. N. Blouin, M. Leclerc, Poly(2,7-carbazole)s: Structure−property relationships. *Acc. Chem. Res.* **41**, 1110–1119 (2008).
40. M. Frisch, G. Trucks, H. Schlegel, G. Scuseria, M. Robb, J. Cheeseman, G. Scalmani, V. Barone, G. Petersson, H. Nakatsuji, Gaussian 16 Revision B. 01, 2016. *Gaussian Inc., Wallingford CT* (2016).
41. L. Wang, G. Nan, X. Yang, Q. Peng, Q. Li, Z. Shuai, Computational methods for design of organic materials with high charge mobility. *Chem. Soc. Rev.* **39**, 423–434 (2010).
42. T. Holstein, Studies of polaron motion: Part II. The "small" polaron. *Ann. Phys.* **8**, 343–389 (1959).
43. L. Wang, D. Beljonne, L. Chen, Q. Shi, Mixed quantum-classical simulations of charge transport in organic materials: Numerical benchmark of the Su-Schrieffer-Heeger model. *J. Chem. Phys.* **134**, 244116 (2011).
44. K. Hannewald, P. A. Bobbert, Ab initio theory of charge-carrier conduction in ultrapure organic crystals. *Appl. Phys. Lett.* **85**, 1535–1537 (2004).
45. L. Wang, D. Beljonne, Flexible surface hopping approach to model the crossover from hopping to band-like transport in organic crystals. *J. Phys. Chem. Lett.* **4**, 1888–1894 (2013).
46. G. Granucci, M. Persico, A. Toniolo, Direct semiclassical simulation of photochemical processes with semiempirical wave functions. *J. Chem. Phys.* **114**, 10608–10615 (2001).
47. F. Plasser, G. Granucci, J. Pittner, M. Barbatti, M. Persico, H. Lischka, Surface hopping dynamics using a locally diabatic formalism: Charge transfer in the ethylene dimer cation and excited state dynamics in the 2-pyridone dimer. *J. Chem. Phys.* **137**, 22A514 (2012).
48. S. Fernandez-Alberti, A. E. Roitberg, T. Nelson, S. Tretiak, Identification of unavoided crossings in nonadiabatic photoexcited dynamics involving multiple electronic states in polyatomic conjugated molecules. *J. Chem. Phys.* **137**, 014512 (2012).
49. L. Wang, D. Trivedi, O. V. Prezhdo, Global flux surface hopping approach for mixed quantum-classical dynamics. *J. Chem. Theory Comput.* **10**, 3598–3605 (2014).
50. M. J. Bedard-Hearn, R. E. Larsen, B. J. Schwartz, Mean-field dynamics with stochastic decoherence (MF-SD): A new algorithm for nonadiabatic mixed quantum/classical molecular-dynamics simulations with nuclear-induced decoherence. *J. Chem. Phys.* **123**, 234106 (2005).
51. A. Troisi, G. Orlandi, Charge-transport regime of crystalline organic semiconductors: Diffusion limited by thermal off-diagonal electronic disorder. *Phys. Rev. Lett.* **96**, 086601 (2006).
52. F. Poulsen, T. Hansen, Band gap energy of gradient core–shell quantum dots. *J. Phys. Chem. C* **121**, 13655–13659 (2017).
53. C. G. Van de Walle, J. Neugebauer, Universal alignment of hydrogen levels in semiconductors, insulators and solutions. *Nature* **423**, 626–628 (2003).
54. S. Adachi, *Handbook on physical properties of semiconductors* (Springer, Boston, MA, 2004), vol. 3.
55. M. Furno, R. Meerheim, S. Hofmann, B. Lüssem, K. Leo, Efficiency and rate of spontaneous emission in organic electroluminescent devices. *Phys. Rev. B* **85**, 115205 (2012).
56. L. E. Brus, A simple-model for the ionization-potential, electron-affinity, and aqueous redox potentials of small semiconductor crystallites. *J. Chem. Phys.* **79**, 5566–5571 (1983).
57. J. Jasieniak, M. Califano, S. E. Watkins, Size-dependent valence and conduction band-edge energies of semiconductor nanocrystals. *ACS Nano* **5**, 5888–5902 (2011).

58. S. Sapra, D. D. Sarma, Evolution of the electronic structure with size in II-VI semiconductor nanocrystals. *Phys. Rev. B* **69**, 125304 (2004).
59. K. Fan, C. Liao, R. Xu, H. Zhang, Y. Cui, J. Zhang, Effect of shell thickness on electrochemical property of wurtzite CdSe/CdS core/shell nanocrystals. *Chem. Phys. Lett.* **633**, 1–5 (2015).
60. J. Park, Y.-H. Won, T. Kim, E. Jang, D. Kim, Electrochemical charging effect on the optical properties of InP/ZnSe/ZnS quantum dots. *Small* **16**, 2003542 (2020).
61. K.-W. Tsai, M.-K. Hung, Y.-H. Mao, S.-A. Chen, Solution-processed thermally activated delayed fluorescent OLED with high EQE as 31% using high triplet energy crosslinkable hole transport materials. *Adv. Funct. Mater.* **29**, 1901025 (2019).
62. D. Di, A. S. Romanov, L. Yang, J. M. Richter, J. P. H. Rivett, S. Jones, T. H. Thomas, M. Abdi Jalebi, R. H. Friend, M. Linnolahti, M. Bochmann, D. Credgington, High-performance light-emitting diodes based on carbene-metal-amides. *Science* **356**, 159–163 (2017).
63. Y. H. Kim, S. Kim, A. Kakekhani, J. Park, J. Park, Y. H. Lee, H. X. Xu, S. Nagane, R. B. Wexler, D. H. Kim, S. H. Jo, L. Martinez-Sarti, P. Tan, A. Sadhanala, G. S. Park, Y. W. Kim, B. Hu, H. J. Bolink, S. Yoo, R. H. Friend, A. M. Rappe, T. W. Lee, Comprehensive defect suppression in perovskite nanocrystals for high-efficiency light-emitting diodes. *Nat. Photon.* **15**, 148–155 (2021).
64. T.-H. Han, M.-R. Choi, C.-W. Jeon, Y.-H. Kim, S.-K. Kwon, T.-W. Lee, Ultrahigh-efficiency solution-processed simplified small-molecule organic light-emitting diodes using universal host materials. *Sci. Adv.* **2**, e1601428 (2016).
65. S. K. Jeon, H.-J. Park, J. Y. Lee, Highly efficient soluble blue delayed fluorescent and hyperfluorescent organic light-emitting diodes by host engineering. *ACS Appl. Mater. Interfaces* **10**, 5700–5705 (2018).
66. Y. Dong, Y.-K. Wang, F. Yuan, A. Johnston, Y. Liu, D. Ma, M.-J. Choi, B. Chen, M. Chekini, S.-W. Baek, L. K. Sagar, J. Fan, Y. Hou, M. Wu, S. Lee, B. Sun, S. Hoogland, R. Quintero-Bermudez, H. Ebe, P. Todorovic, F. Dinic, P. Li, H. T. Kung, M. I. Saidaminov, E. Kumacheva, E. Spiecker, L.-S. Liao, O. Voznyy, Z.-H. Lu, E. H. Sargent, Bipolar-shell resurfacing for blue LEDs based on strongly confined perovskite quantum dots. *Nat. Nanotechnol.* **15**, 668–674 (2020).
67. H. Li, H. Li, Y. Zhi, J. Wang, L. Tang, Y. Tao, G. Xie, C. Zheng, W. Huang, R. Chen, Multiple σ–π conjugated molecules with selectively enhanced electrical performance for efficient solution-processed blue electrophosphorescence. *Adv. Opt. Mater.* **7**, 1901124 (2019).
68. X. Li, Q. Lin, J. Song, H. Shen, H. Zhang, L. S. Li, X. Li, Z. Du, Quantum-dot light-emitting diodes for outdoor displays with high stability at high brightness. *Adv. Opt. Mater.* **8**, 1901145 (2020).
69. K. Lin, J. Xing, L. N. Quan, F. P. G. de Arquer, X. Gong, J. Lu, L. Xie, W. Zhao, D. Zhang, C. Yan, W. Li, X. Liu, Y. Lu, J. Kirman, E. H. Sargent, Q. Xiong, Z. Wei, Perovskite light-emitting diodes with external quantum efficiency exceeding 20 per cent. *Nature* **562**, 245–248 (2018).
70. L. Zhang, F. Yuan, J. Xi, B. Jiao, H. Dong, J. Li, Z. Wu, Suppressing ion migration enables stable perovskite light-emitting diodes with all-inorganic strategy. *Adv. Funct. Mater.* **30**, 2001834 (2020).
71. Y.-F. Chang, C.-H. Yu, S.-C. Yang, I. H. Hong, S.-C. Jiang, H.-F. Meng, H.-L. Huang, H.-W. Zan, S.-F. Horng, Great improvement of operation-lifetime for all-solution OLEDs with mixed hosts by blade coating. *Org. Electron.* **42**, 75–86 (2017).

72. Y. Suzuki, Q. Zhang, C. Adachi, A solution-processable host material of 1,3-bis{3-[3-(9-carbazolyl)phenyl]-9-carbazolyl}benzene and its application in organic light-emitting diodes employing thermally activated delayed fluorescence. *J. Mater. Chem. C* **3**, 1700–1706 (2015).
73. L. Wang, J. Lin, Y. Hu, X. Guo, Y. Lv, Z. Tang, J. Zhao, Y. Fan, N. Zhang, Y. Wang, X. Liu, Blue quantum dot light-emitting diodes with high electroluminescent efficiency. *ACS Appl. Mater. Interfaces* **9**, 38755–38760 (2017).
74. C. Bi, Z. Yao, X. Sun, X. Wei, J. Wang, J. Tian, Perovskite quantum dots with ultralow trap density by acid etching-driven ligand exchange for high luminance and stable pure-blue light-emitting diodes. *Adv. Mater.* **33**, 2006722 (2021).

**Competing interests:** The authors declare no competing interests.

**Data and materials availability:** All data in the main text or the supplementary materials are available on request from the corresponding authors.

**Supplementary Materials**

Materials and Methods

Supplementary Text

Figs. S1 to S13

Tables S1 to S2

References (*40–74*)

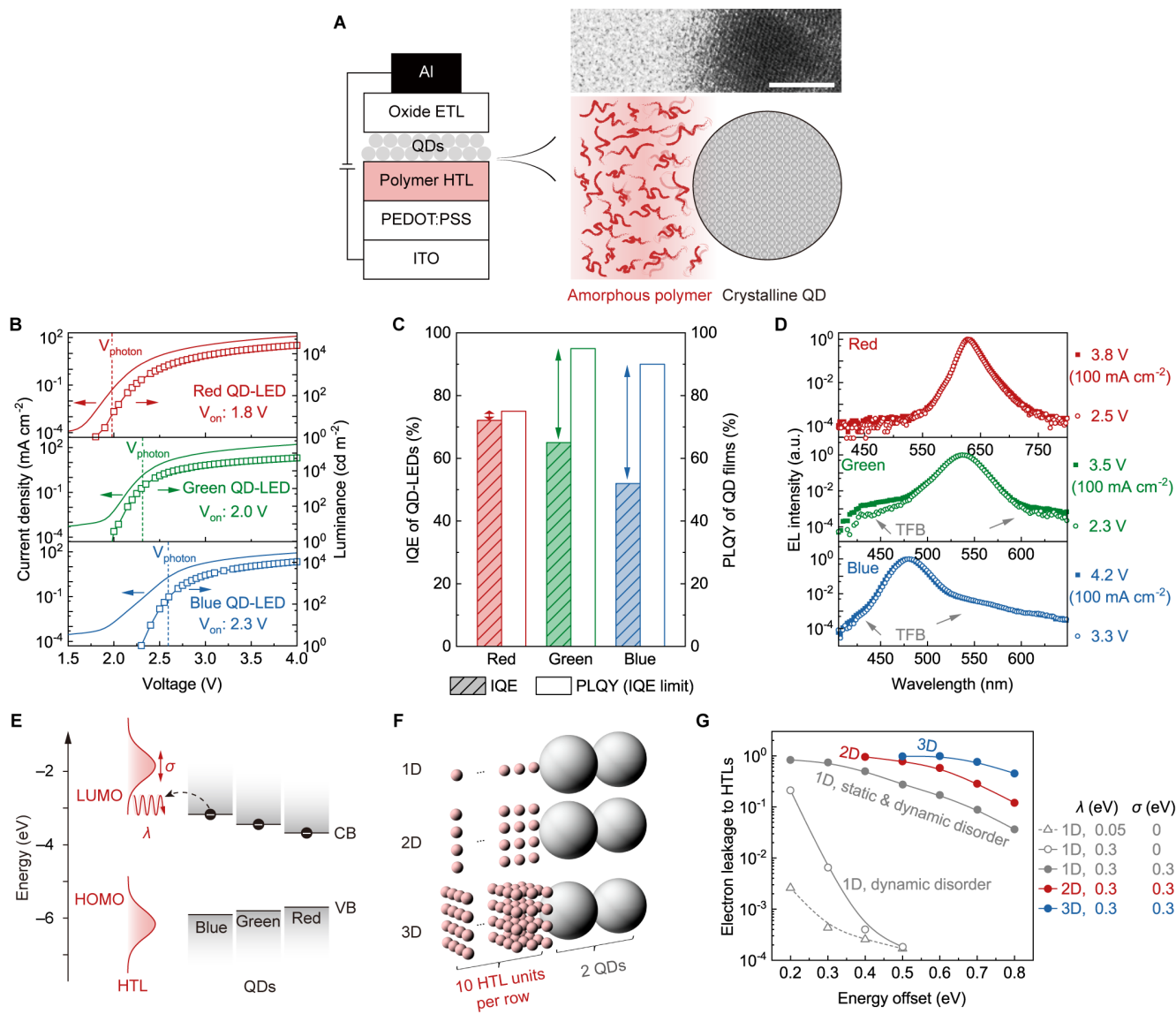

**Fig. 1. Disorder-enhanced electron leakage limits EL efficiencies of blue and green QD-LEDs.** (**A**) Typical QD-LED structure (left), HTL/QD interface (bottom right) and a cross-sectional TEM image of the interface (top right, scale bar: 5 nm). (**B**) Current density-voltage-luminance characteristics of the QD-LEDs. (**C**) EL-PL efficiency gaps of the QD-LEDs. (**D**) EL spectra of the QD-LEDs (semi-log scale). (**E**) Electron transfer from QDs to polymeric HTLs modulated by energetic disorders of HTLs. (**F**) Schematic representation of the 1D, 2D, and 3D models for the interfacial electron-transfer simulations. (**G**) Impacts of energetic disorders and geometric factors on the electron-leakage probabilities.

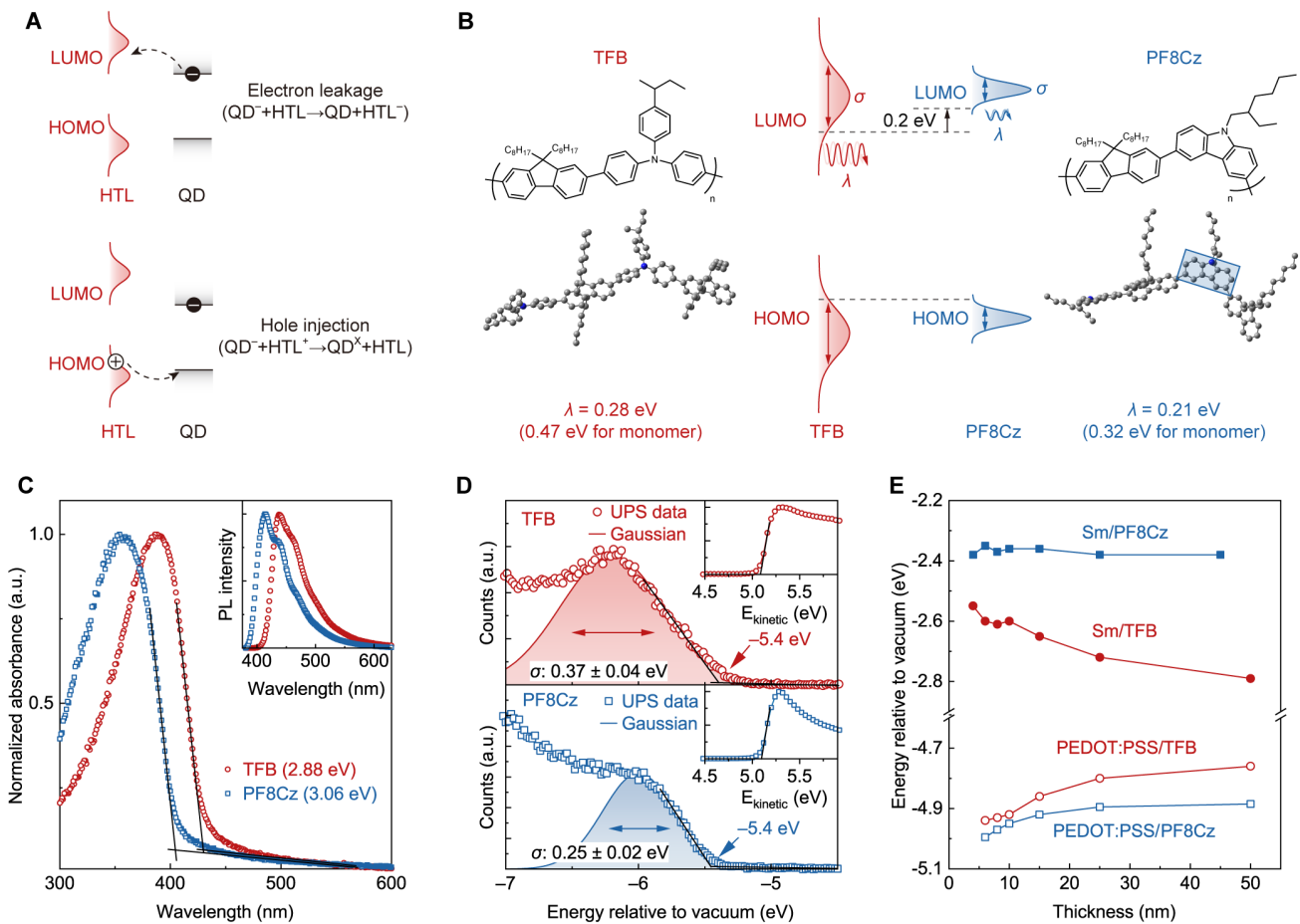

**Fig. 2. HTL design to eliminate the electron-leakage channel.** (**A**) Two competitive processes of the negatively charged QDs at the HTL/QD interface: electron leakage (top) and hole injection (bottom). (**B**) Chemical structures, theoretically optimized geometries, and reorganization energies ($\lambda$) for TFB and PF8Cz dimers. Middle: A comparison of the electronic structures. (**C**) Absorption and PL spectra of the HTLs. (**D**) UPS results of the valence spectra and the corresponding secondary-electron edges (insets). Gaussian fits of the high-kinetic-energy edges and the fitted widths ($\sigma$, instrumental broadening decoupled) are also shown. (**E**) The thickness-dependent surface work function of the HTLs on the substrates of Sm or PEDOT:PSS.

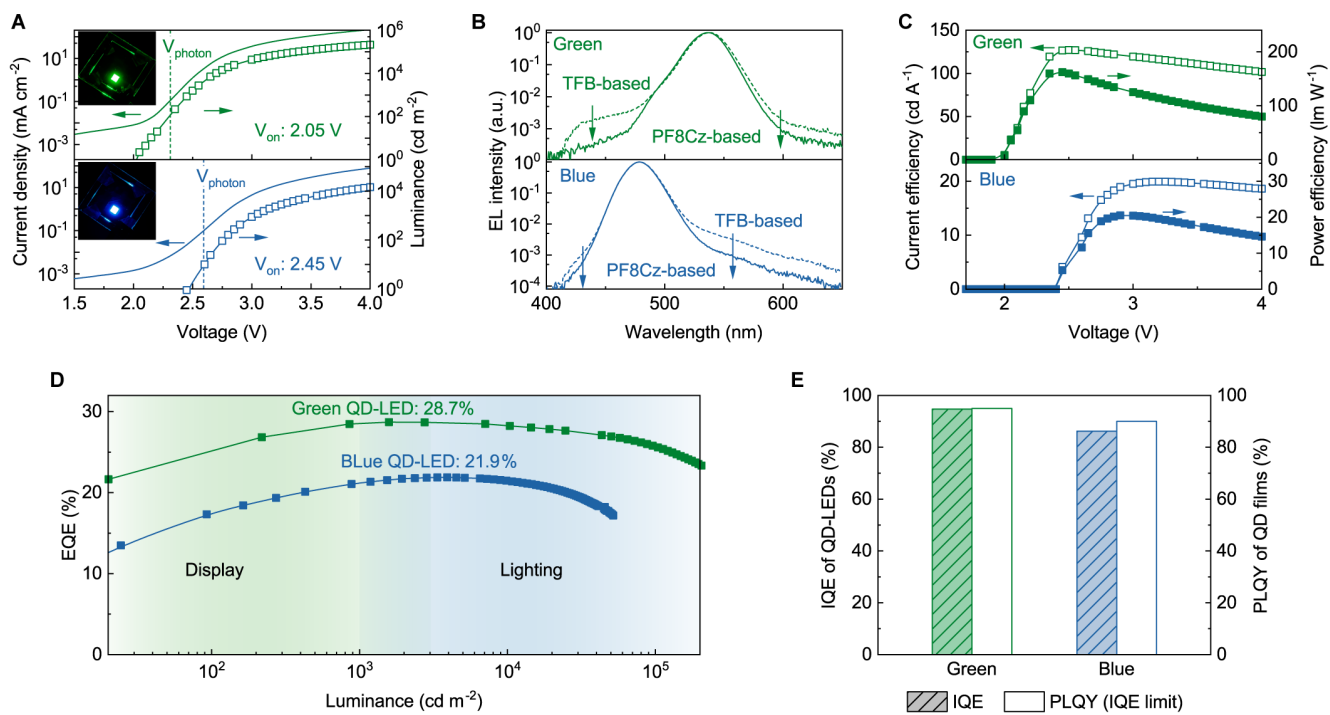

**Fig. 3. High-efficiency green and blue QD-LEDs based on the PF8Cz HTLs.** (**A**) Current density-voltage-luminance characteristics of the green (top) and blue (bottom) QD-LEDs. (**B**) EL spectra of the green and blue QD-LEDs based on the PF8Cz HTLs (solid lines) and those of the devices based on the TFB HTLs (dashed lines). The arrows indicate the decrease of the parasitic emissions from HTLs in the PF8Cz-based devices compared with those in the TFB-based devices. (**C**) Current efficiencies (left axis) and power efficiencies (right axis) of the green and blue QD-LEDs. (**D**) EQEs of the green and blue QD-LEDs in the luminance range for display and general lighting applications. (**E**) IQEs of the green and blue QD-LEDs and the corresponding PLQYs of the QD films.

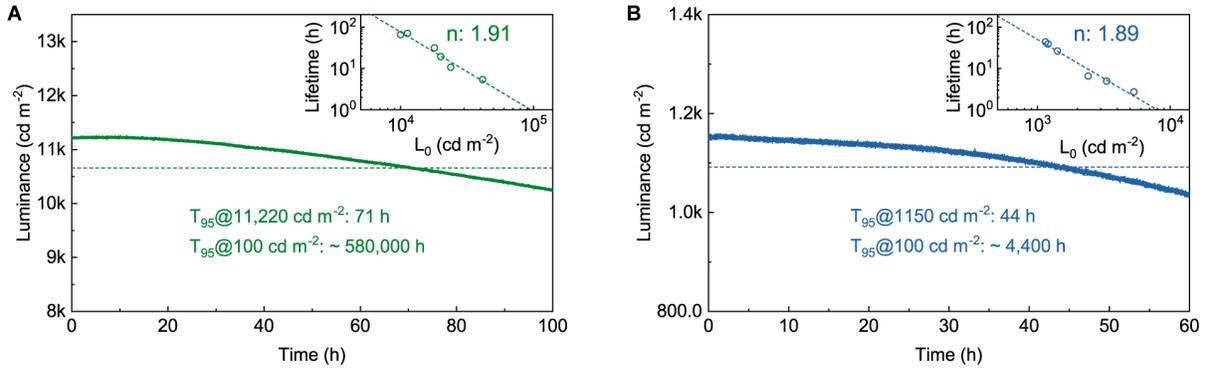

**Fig. 4. Operational lifetimes of the QD-LEDs based on the PF8Cz HTLs.** Luminance as a function of operational time for a green (**A**) and a blue (**B**) QD-LED. The lifetimes ($T_{95}$) at various initial luminance ($L_0$) are shown in the insets. The acceleration factors (n) are fitted according to the empirical relationship of $L_0^n T_{95}$ = constant.